\documentclass{PoS}
\usepackage{wrapfig,amsmath,amssymb,MnSymbol}

\DeclareMathOperator{\Tr}{Tr}

\def\beann{\begin{eqnarray*}} 
\def\eeann{\end{eqnarray*}}

\def\T{{\cal T}}

\title{The Hubbard model in the canonical formulation}

\ShortTitle{The Hubbard model in the canonical formulation}

\author{Sebastian Burri, \speaker{Urs Wenger}\\
       Albert Einstein Center for Fundamental Physics\\
        Institute for Theoretical Physics\\
        University of Bern\\
        Sidlerstrasse 5\\
        CH--3012 Bern\\
        Switzerland\\
        E-mail: \email{burri@itp.unibe.ch}, \email{wenger@itp.unibe.ch}}

      \abstract{We describe non-relativistic fermions on
        the lattice (Hubbard model) in the canonical formulation using
        transfer matrices in fixed fermion number sectors such that
        the partition function becomes fully factorized in time. By
        analytically integrating out the auxiliary
        Hubbard-Stratanovich field due to the four-fermion
        interaction, we express the system in terms of discrete, local
        fermion occupation numbers which are the only remaining
        degrees of freedom. We show the close relation to the fermion
        loop and the fermion bag formulation. One can prove that in
        1+1 dimension the fermion sign problem is absent. Finally, we
        construct improved estimators for the ground state energy,
        $2$-point functions, and for the chemical
        potential.}

\FullConference{37th International Symposium on Lattice Field Theory - Lattice2019\\
		16-22 June 2019\\
		Wuhan, China}

\begin{document}

\section{Motivation}
The sign problem at finite fermion density can be understood as a
manifestation of huge cancellations between different states which
contribute to the partition function in a given basis. Obviously, all
states are always present in the grand-canonical partition function
$Z(\mu,T)$ for any values of the chemical potential $\mu$ and
temperature $T$. However, only some states are relevant and important
for the physics at specific values of $\mu$ and $T$ and contribute
significantly to $Z(\mu,T)$, while all other states either have
negligible weights or need to cancel out. A prominent example is given
by the so-called Silver Blaze property of physical systems: certain
physical states are completely invisible in the physical properties of
the system up to the critical chemical potential $\mu_c$, when they
suddenly become the main physical degrees of freedom driving the
physics.

In the canonical formulation, fewer cancellations are necessary to
describe the same physics, as compared to the grand-canonical
formulation. This becomes clear by considering the dimension of the
Fock space.  Consider for example a system specified by the
Hamiltonian ${\cal H}(\mu)$ and its grand-canonical partition function
\[
Z_\textrm{GC}(\mu,T) = \Tr \left[e^{-{\cal H}(\mu)/T}\right]  = \Tr \prod_t \T_t(\mu)\, , 
\]
where in the last step we assumed a discretized time and a Trotter
decomposition, such that $\T_t(\mu)$ are transfer matrices at fixed
time indices $t$.  In the canonical formulation, the partition
function reads
\[
Z_\text{C}(N_f,T) = \Tr_{N_f}\big[  e^{-{\cal H}/T}\big]  = \Tr \prod_t
\T^{(N_f)}_t \, ,
\]
where the trace is now restricted to states with a specific fermion
number $N_f$.  As a consequence, in the canonical formulation certain
cancellations are more obvious than in the grand-canonical one. In
QCD, for example, the canonical partition functions vanish for the
number of quarks $N_Q \neq 0$ mod $3$ due to the transformation
properties of the Fock states, while in the grand-canonical
formulation the cancellation is not obvious at all. Similarly, in the
$N_f=1$ Schwinger model it is immediately obvious that in the
canonical formulation only the states in the zero-charge sector are
physical, because the contributions from the canonical states in the
other sectors cancel due to the global U(1) symmetry, while in the
grand-canonical formulation the cancellation is not explicit.

The canonical transfer matrices can be constructed for a large class
of quantum field theories.  In some supersymmetric Yang-Mills gauge
theories, for example, the canonical formulation has been shown to
lead to a solution of the fermion sign problem
\cite{Bergner:2015ywa,Bergner:2016qbz}, based on the construction of
the transfer matrices in and the close connection of the canonical
formulation with the (dual) fermion loop or worldline formulation
\cite{Steinhauer:2014oda}. This, in turn, allows for the construction
of efficient simulation algorithms, such as fermion bag
\cite{Chandrasekharan:2009wc,Chandrasekharan:2013rpa} and fermion worm
algorithms \cite{Wenger:2008tq}.  Based on the construction of the
canonical transfer matrices in QCD \cite{Alexandru:2010yb}, one can
show that the sign problem is absent in the strong coupling and
heavy-dense limit, even though it is very severe just away from
it. Nevertheless, in the related 3-state Potts model the sign problem
at finite density can be solved by defining clusters with fixed
fermion numbers in the canonical formulation
\cite{Alexandru:2017dcw}. In this context, it is worthwhile to note
that the fermionic degrees of freedom in the canonical formulation are
expressed in terms of local fermion occupation numbers $n_x =0,1$, and
the only fermionic degrees of freedom are discrete index sets.

One can also construct improved estimators for fermionic correlation
functions using the canonical transfer matrices, potentially allowing
for local (multi-level) update schemes with exponential improvement of
the signal-to-noise ratio, or even for the direct Monte-Carlo sampling
of the correlation function.  In some cases, the canonical formulation
makes it possible to integrate out explicitly the (auxiliary) bosonic
degrees of freedom, such as the Hubbard-Stratanovich field in the
Hubbard model, thereby solving the fermion sign problem explicitly in
$d=1+1$ dimension.

In these proceedings, we consider the specific example of the Hubbard
model and demonstrate in detail the connections and constructions
mentioned above. We note that some of these ideas have been considered
in the Hubbard model some time ago in
\cite{PhysRevD.24.2278,Hirsch:1983wk,PhysRevB.34.3216,PhysRevB.40.506}.

\section{Canonical formulation of the Hubbard model}
Let us consider the Hamiltonian for the Hubbard model,
\[ 
{\cal H}(\mu) = - \sum_{\langle x,y\rangle, \sigma} t_\sigma \,
\hat c_{x,\sigma}^\dagger \hat c_{y,\sigma} +\sum_{x,\sigma}
\mu_\sigma {N}_{x,\sigma} + U \sum_x
{N}_{x,\uparrow}{N}_{x,\downarrow} \, ,
\]
where $\sigma=\downarrow,\uparrow$ denotes two fermion species, e.g.,
with spin up or down, $c_{x,\sigma}^\dagger$ and $c_{x,\sigma}$ are
the creation and annihilation operators for the corresponding fermion
at position $x$, while $\langle x,y\rangle$ denotes the nearest
neighbours and $t_\sigma$ the hopping parameter. The four-fermion
interaction between the fermions is parametrized by the local particle
number ${N}_{x,\sigma} = \hat c_{x,\sigma}^\dagger \hat c_{x,\sigma}$
and the coupling strength $U$, while the total particle number is
tuned through the chemical potentials $\mu_\sigma$.

The grand-canonical partition function can now be written as
\[
Z_\text{GC}(\mu) = \Tr \left[e^{-{\cal H}(\mu)/T} \right] 
                = \sum_{\{{N}_\sigma\}} e^{-\sum_\sigma {N}_\sigma
                  \mu_\sigma/T } \cdot \, Z_C(\{{N}_\sigma\}) \, ,
\]
where in the last step we perform a fugacity expansion in order to
express the grand-canonical partition function in terms of the
canonical ones, i.e., $Z_C(\{{N}_\sigma\})=\Tr \prod_t {\cal
  T}_t^{(\{{N}_\sigma\})}$. Note that the canonical transfer matrices
are now restricted to states with fixed total fermion number $N_\sigma
= \sum_x {N}_{x,\sigma}$.

In order to investigate the model nonperturbatively using Monte Carlo
simulations, one usually performs a Trotter decomposition and
introduces a coherent state representation for the fermionic degrees
of freedom, eventually yielding the partition function in terms of a
path integral over Grassmann-valued fields $\psi^\dagger$ and $\psi$,
\[
Z_\text{GC}(\mu) = \int {\cal D}\psi^\dagger {\cal D}\psi
e^{-S[\psi^\dagger,\psi;\mu]} \, ,
\] 
with Euclidean action
\[
S[\psi^\dagger, \psi;\mu] = \sum_\sigma
\psi_{\sigma}^\dagger \nabla_t \psi_{\sigma} + H[\psi^\dagger,\psi;\mu]\, ,
\]
where the Hamilton function $H$ describes the spatial hoppings of the
Grassmann fields as well as the four-fermion interaction. This
interaction can be bilinearized by a Hubbard-Stratonovich (HS)
transformation using the HS field $\phi$ leading to
\[
Z_\text{GC}(\mu) = \int {\cal D}\psi^\dagger {\cal D}\psi {\cal D}
\phi \, \rho[\phi] \, e^{- \sum_\sigma
  S[\psi^\dagger_\sigma,\psi_\sigma,\phi;\mu_\sigma]} \, ,
\]
with the action $S[\psi^\dagger_\sigma,\psi_{\sigma},\phi;\mu_\sigma]
= \psi^\dagger_\sigma M[\phi;\mu_\sigma] \psi_{\sigma}$ now being a
bilinear form in the fermion fields $\psi^\dagger_\sigma,\psi_\sigma$
involving the fermion matrix $M$, and $\rho[\phi]$ being the Boltzmann
factor for the HS field $\phi$. The fermion fields can now be
integrated out, yielding the determinant of $M$,
\[
Z_\text{GC}(\mu) = \int {\cal D} \phi \, \rho[\phi] \prod_\sigma \det M[\phi;\mu_\sigma] \, .
\]
The HS transformation can be chosen in such a way,
cf.~e.g.~\cite{Alexandru:2018brw}, that the fermion matrix has the
structure
\[
M[\phi;\mu_\sigma] = \left(
\begin{array}{cclc}
B & 0 & \ldots & \pm e^{\mu_\sigma} C(\phi_{L_t-1}) \\
-e^{\mu_\sigma} C(\phi_0) & B & \ldots & 0 \\
\vdots & \ddots & \ddots & \vdots \\
0 & \ldots & -e^{\mu_\sigma} C(\phi_{L_t-2}) & B
\end{array} \right)
\]
where $B$ and $C$ are matrices of size $L_s\times L_s$, and $L_s$ and
$L_t$ are the number of spatial and temporal lattice points. Note that
while $B$ is a constant matrix, $C = C(\phi_t)$ only depends on the HS
fields on time slice $t$ and can be chosen diagonal.  The determinant
of $M$ can be dimensionally reduced to \cite{PhysRevD.24.2278}
\[
\det M[\phi;\mu_\sigma] = \det B^{L_t} \cdot \det\left(1 \mp e^{L_t \mu_\sigma} {\cal T}[\phi]\right)
\]
where ${\cal T}[\phi] = B^{-1} C(\phi_{L_t-1})\cdot \ldots \cdot
B^{-1} C(\phi_0)$. The fugacity expansion of this object yields the
canonical determinants $\det M_{{N}_\sigma}[\phi]$ as the coefficients
of the corresponding fugacity terms. They can be expressed as a sum
over the principal minors $\det {\cal T}^{\, \backslash
  \hspace{-0.15cm} J \, \backslash \hspace{-0.15cm}J}$ of order
${N}_\sigma$,
\begin{equation}\label{eq:canonical determinant}
\det M_{{N}_\sigma}[\phi]= \sum_J \det {\cal T}^{\, \backslash \hspace{-0.15cm} J  \, \backslash  \hspace{-0.15cm}J}[\phi]  = \Tr \left[\prod_t {\cal T}_t^{({N}_\sigma)}[\phi_t]\right] \, ,
\end{equation}
where the index sets $J \subset \{1,\ldots,L_s\}$ of size ${N}_\sigma
= |J|$ essentially label the states of the fermion Fock space and
therefore allow to identify the sum as the trace over the product of
transfer matrices acting in the Fock space of states with fixed
fermion number ${N}_\sigma$. Indeed, in sector ${N}_\sigma$ the number
of states $N_\text{states} = \binom{L_s}{{N}_\sigma}$ is equal to the
number of principal minors of order ${N}_\sigma$.  At half-filling,
this number grows exponentially with $L_s$, however, it can
efficiently be evaluated stochastically with Monte Carlo methods, as
for example in \cite{Bergner:2015ywa,Bergner:2016qbz}.

\section{Transfer matrices of the Hubbard model}
Using eq.~(\ref{eq:canonical determinant}) we can now identify the
transfer matrices and write down explicit expressions for them. To
this end we recall the Cauchy-Binet formula
\[
\det(A\cdot B)^{\, \backslash \hspace{-0.15cm} I  \, \backslash
  \hspace{-0.15cm}K} = \det A^{\, \backslash \hspace{-0.15cm} I  \,
  \backslash  \hspace{-0.15cm}J} \cdot \det B^{\, \backslash \hspace{-0.15cm} J  \, \backslash  \hspace{-0.15cm}K}
\]
and factorize ${\cal T}[\phi]$ into the product of transfer matrices
${\cal T}_t^{({N}_\sigma)}[\phi_t]$, which are hence given by
\begin{equation*}
(\T_t^{({N}_\sigma)})_{IK} = \det B \cdot \det \left[B^{-1} \cdot C(\phi_t)\right]^{ \,
  \backslash \hspace{-0.15cm} I  \, \backslash  \hspace{-0.15cm}K}
 = \det B \cdot \det (B^{-1})^{ \, \backslash \hspace{-0.15cm} I  \,
  \backslash  \hspace{-0.15cm}J} \cdot \det C(\phi_t)^{ \, \backslash
  \hspace{-0.15cm} J  \, \backslash  \hspace{-0.15cm}K} \, ,
\end{equation*}
where $|I|=|J|=|K|=N_\sigma$.  Moreover, using the complementary
cofactor we have
\[
\det B \cdot \det (B^{-1})^{ \, \backslash \hspace{-0.15cm} J  \,
  \backslash  \hspace{-0.15cm}I} = (-1)^{p(I,J)} \det B^{IJ}\,,
\]
where {\small $p(I,J) = \sum_i(I_i+J_i)$}.  Since the HS
transformation can be chosen such that $C(\phi_t)$ is diagonal, we
find
\[
\det C(\phi_t)^{ \, \backslash \hspace{-0.15cm} J  \, \backslash
  \hspace{-0.15cm}K} = \delta_{JK} \prod_{x \notin J} \phi_{x,t}
\]
and the HS field can be integrated out site by site,
\[
\int d\phi_{x,t} \, \rho(\phi_{x,t}) \,
\phi_{x,t}^{\sum_\sigma\delta_{x\notin  J^\sigma}} 
\equiv w_{x,t} 
=\left\{ \begin{array}{l}
w_2, \quad \text{if } x \notin J^\uparrow, x \notin J^\downarrow, \\
w_1, \quad \text{else, } \\
w_0, \quad \text{if } x \in J^\uparrow, x \in J^\downarrow,
\end{array}
\right.
\]
where $w_i > 0$.  Finally, with $\prod_x w_{x,t} \equiv W
\left(\{J_t^\sigma\}\right)$ we have the canonical partition function
of the Hubbard model expressed in fully factorized form,
\begin{equation}\label{eq:factorized canonical partition function}
Z_C(\{{N}_\sigma\}) = \sum_{\{J_t^\sigma\}} \prod_t \left(\prod_\sigma \det B^{J_{t-1}^\sigma
  J_{t}^\sigma} \right) W 
\left(\{J_t^\sigma\}\right), \quad |J_t^\sigma|={N}_\sigma \, ,
\end{equation}
with the only dynamical degrees of
freedom being the index sets $\{J_t^\sigma\}$ on each time slice.
\begin{wrapfigure}{r}{0.5\textwidth}
\vspace*{-0.3cm}
\begin{minipage}[b]{5.7cm}
\includegraphics[width=5.5cm]{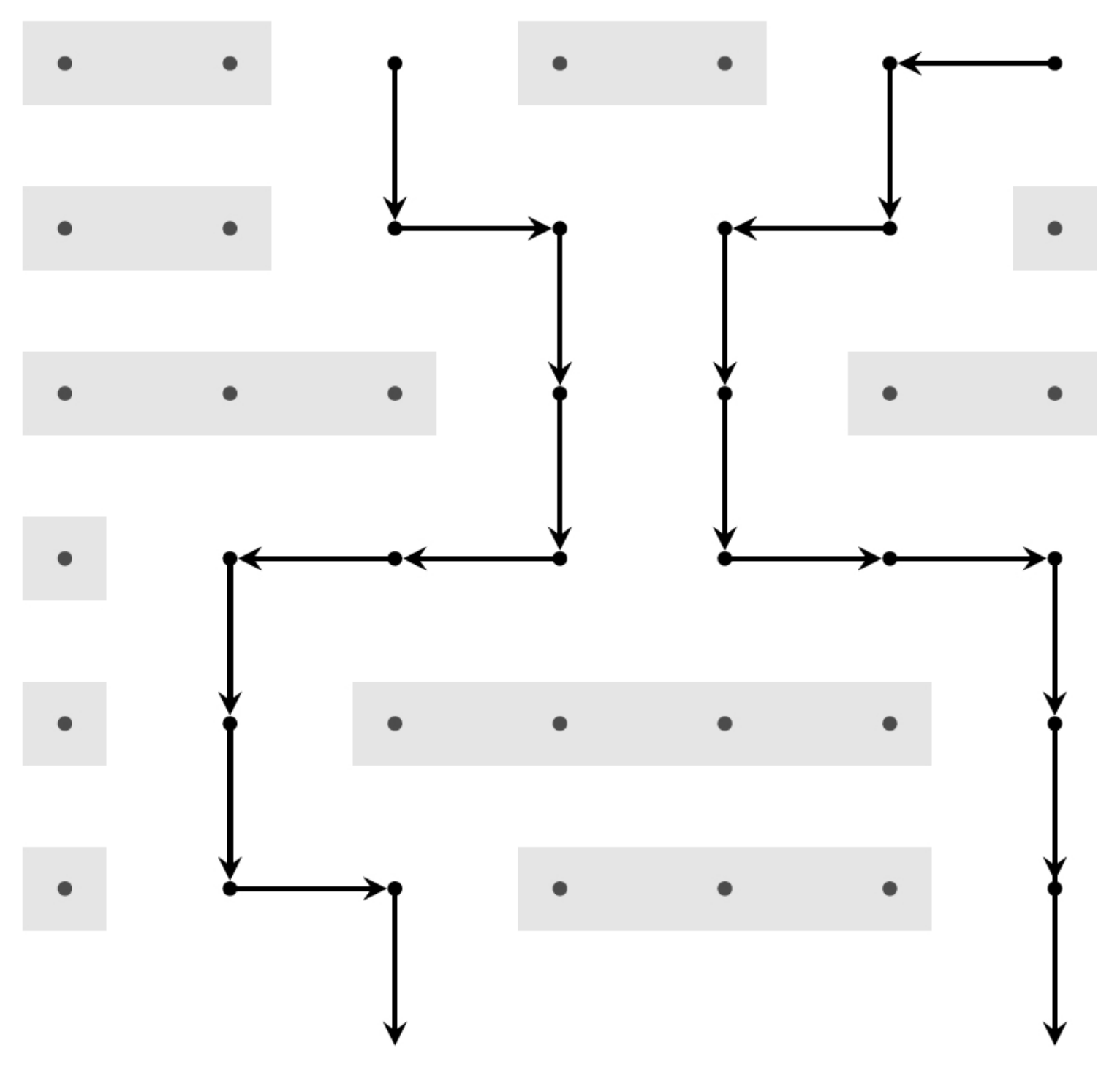}
\end{minipage}
\begin{minipage}[b]{0.5cm}
{\small
\hspace*{0.3cm}$J_t$:\\
\vspace*{-0.25cm}
 \{3,6\}\\
\vspace*{-0.25cm}
 \{4,5\}\\
\vspace*{-0.25cm}
 \{4,5\}\\
\vspace*{-0.25cm}
 \{2,7\}\\
\vspace*{-0.25cm}
 \{2,7\}\\
\vspace*{-0.25cm}
 \{3,7\}\\
\vspace*{-0.1cm}
}
\end{minipage}
\caption{Illustration of a fermion loop configuration
and the corresponding index sets $J_t$.
 \label{fig:fermion loop illustration}}
\vspace{-0.2cm} 
\end{wrapfigure}
The representation in eq.~(\ref{eq:factorized canonical partition
  function}) exposes the relation to the fermion loop
formulation. Interpreting the spatial indices in the index sets
$J_t^\sigma$ as labelling the positions of the $N_\sigma$ fermions
where they hop in time, one can identify the $N_\sigma$ fermion loops
running around the lattice in temporal direction, cf.~Figure
\ref{fig:fermion loop illustration} for an illustration with
$N_\sigma=2$ in $d=1$ spatial dimension. The factors $\det
B^{J_{t-1}^\sigma J_{t}^\sigma}$ collect all the contributions from
different fermion loop configurations on time slice $t$ within the
shaded regions. They can therefore be interpreted as ``fermion
bags''. It turns out that in $d=1$ spatial dimensions the
contributions $\det B^{IJ}$ can be calculated analytically through the
recursion relation \cite{Endres:2012vd}
\[
{\includegraphics[width=8.0cm]{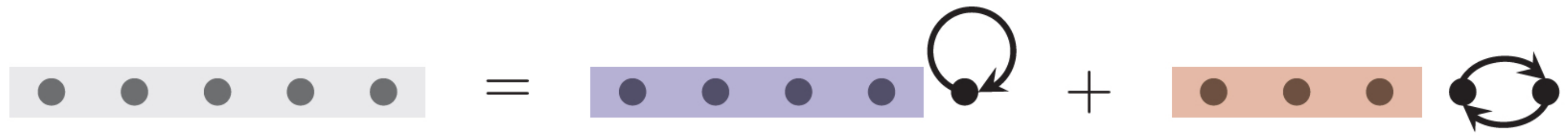}}
\]
and for open b.c.~one can prove that $\det B^{IJ} \geq 0$, i.e., there
is no sign problem. For periodic b.c.~there is no sign problem either,
because in the thermodynamic limit, the partition functions become
independent of the boundary conditions.

\section{Improved estimators}
\begin{figure}[t!]
\vspace*{-1.0cm}
\hspace*{-0.5cm}
\includegraphics[width=8.5cm]{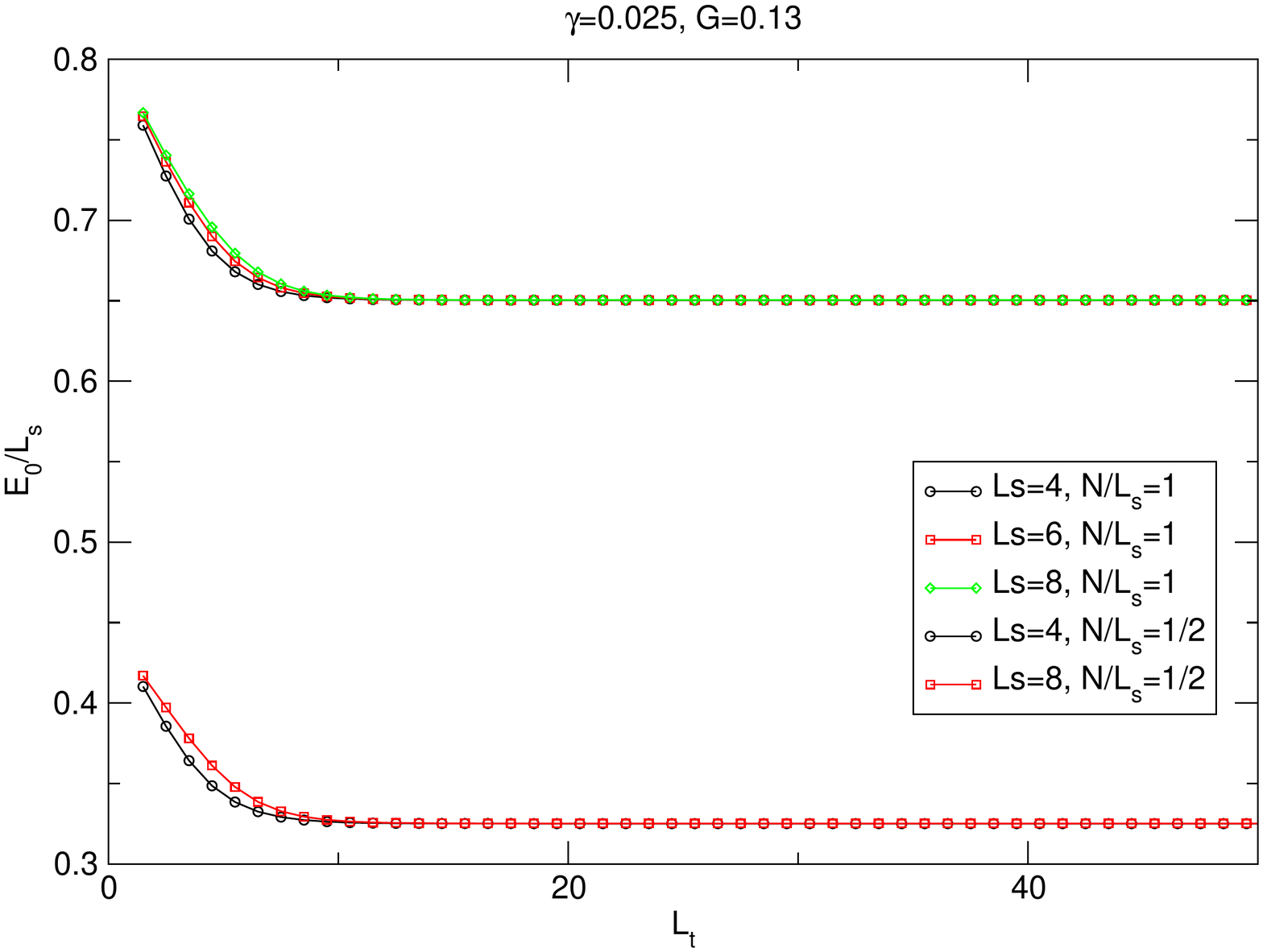} 
\hspace{-1cm}
\includegraphics[width=8.5cm]{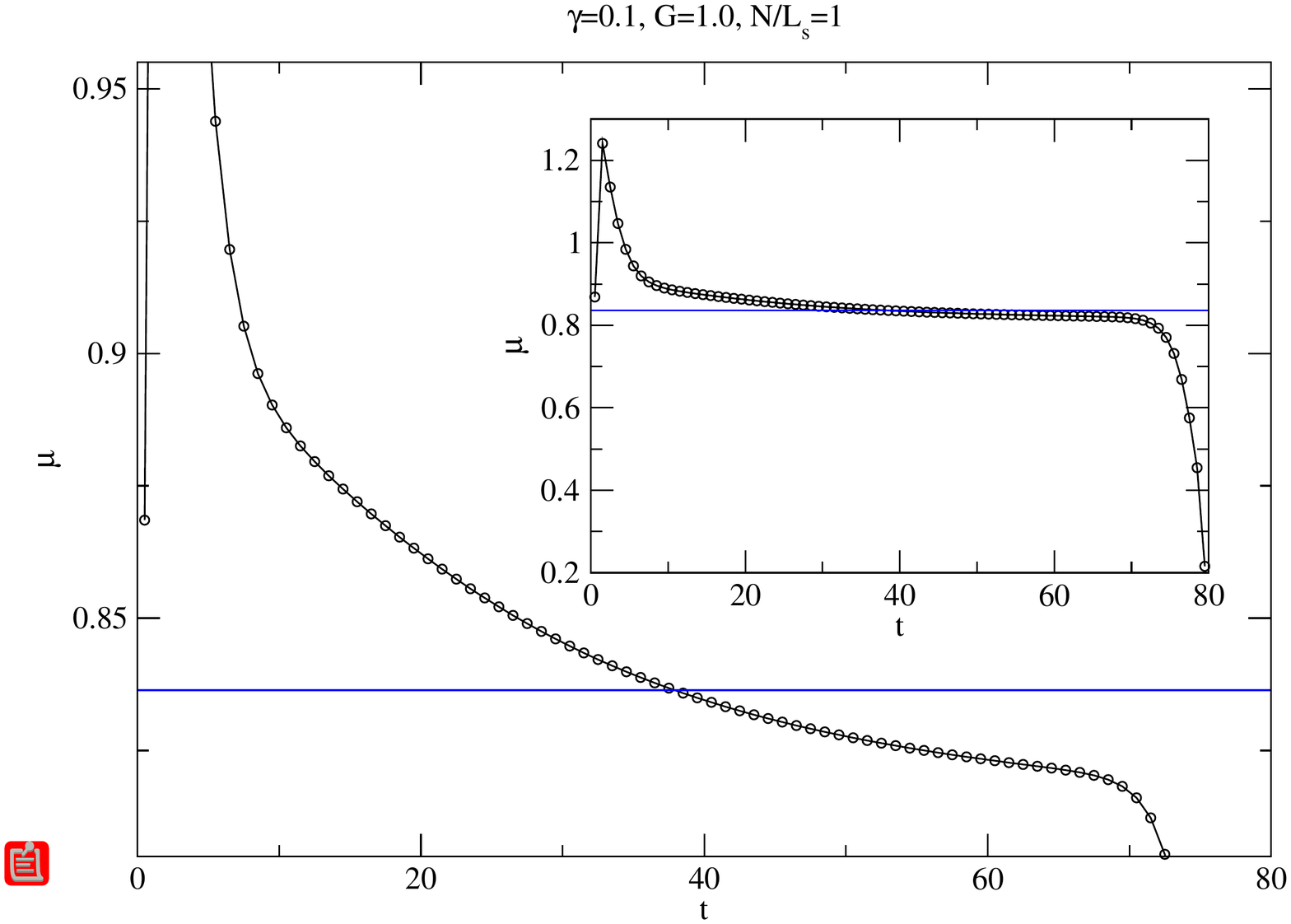}
\vspace*{-1.0cm}
\caption{{\it Left plot:} Logarithm of the partition function ratios
  $Z_C(L_t)/Z_C(L_t+1)$ as a function of $L_t$ which yields the ground
  state energy $E_0$ in the limit $L_t\rightarrow \infty$.  {\it Right
    plot:} Chemical potential from eq.~(\ref{eq:ratios of 2pt-fct}) as
  a function of $t$. The blue line is the value obtained via
  eq.~(\ref{eq:Zratio}) from the ratio $Z_C(n,n_s)/Z_C(n+1,n_s+1)$.
  \label{fig:E0 and mu}}
\end{figure}

Having the canonical partition function in its fully factorized form,
we can construct improved estimators for various observables.  The
energy of the ground state, for example, is defined by $E_0 =
-\lim_{\beta \rightarrow \infty} \partial_\beta \ln Z_C(\beta)$,
which on the lattice becomes
\[
E_0 = \lim_{L_t\rightarrow \infty} 
\ln \frac{Z_C(L_{t})}{Z_C(L_{t}+1)} \, .
\]
Since our formulation is factorized in time, we can write the ratio as
an observable calculated in the ensemble defining $Z_C(L_{t}+1)$,
\[
\frac{Z_C(L_{t})}{Z_C(L_{t}+1)} = \left\langle
\prod_{\sigma}\left(\frac{\det B^{J^\sigma_{t-1}J^\sigma_{t+1}}}{\det
    B^{J^\sigma_{t-1}J^\sigma_t} \det
    B^{J^\sigma_{t}J^\sigma_{t+1}}}\right) \frac{1}{W(\{J^\sigma_t\})}
\right\rangle_{Z_C(L_{t}+1)} \, .
\]
In the left plot of Figure \ref{fig:E0 and mu} we show the logarithm
of this ratio (normalized by the volume $L_s$) as a function of the
inverse temperature $1/T=L_t$ for two different fermion densities
$\rho = N/L_s = (N_\uparrow+N_\downarrow)/L_s=1$ (half-filling) and
$1/2$ (quarter-filling) for specific values of the hopping
$t_\uparrow=t_\downarrow$ and the four-fermion coupling $U$
(parametrized by $\gamma$ and $G$).
As another example we consider the chemical potentials $\mu_{(s)}$
defined as $ \mu_{(s)} = \partial_{\rho_{(s)}} F(\rho_{(s)})$ with
$\rho_{(s)} = ({N}_\uparrow \pm {N}_\downarrow)/L_s$ which on the
lattice can be written as, e.g.,
\begin{equation}\label{eq:Zratio}
\mu = \frac{F(n+2,n_s)-F(n,n_s)}{2} = -\frac{1}{2 L_t} \ln \left( 
\frac{Z_C(n+2,n_s)}{Z_C(n,n_s)}\right) \, .
\end{equation}
Now we define the partition function for the fermionic 2-pt.~function
of, e.g., an up-spin fermion in the background of $\{N_\sigma\}$
fermions,
\[
Z_t^{\text{2-pt.},\uparrow}
(\{N_\sigma\})
= \Tr \left[ P^\dagger \prod_{t'<t}
  \T_{t'}^{\{N_\uparrow+1,N_\downarrow\}} P \cdot \prod_{t'\geq t}
  \T_{t'}^{\{N_\sigma\}} \right] 
= \llangle \psi^\dagger_0 \psi_t \rrangle_{\{N_\sigma\}} \,,
\]
where $P^\dagger$ ($P$) create (annihilate) an up-spin fermion at time
$t'=0$ ($t'=t$). Then we can write the chemical potential as a
telescopic product of ratios of 2-pt.~functions,
\begin{align*}
e^{-2\mu/T}
&= \frac{Z_0^{\text{2-pt.},\uparrow}}{Z(n,n_s)} \cdot
   \frac{Z_1^{\text{2-pt.},\uparrow}}{Z_0^{\text{2-pt.},\uparrow}} \cdot
 \cdot  
\ldots \cdot
 \frac{Z(n+1,n_s+1)}{Z_{Lt-1}^{\text{2-pt.},\uparrow}} 
\times
\frac{Z_0^{\text{2-pt.},\downarrow}}{Z(n+1,n_s+1)} \cdot 
\frac{Z_1^{\text{2-pt.},\downarrow}}{Z_0^{\text{2-pt.},\downarrow}}
\cdot \ldots \cdot
\frac{Z(n+2,n_s)}{Z_{L_t-1}^{\text{2-pt.},\downarrow}} \, .
\end{align*}
Each ratio can in turn be written as an expectation value, e.g.,
\begin{equation} \label{eq:ratios of 2pt-fct}
\frac{Z_{t+1}^{\text{2-pt.},\uparrow}}{Z_{t}^{\text{2-pt.},\uparrow}} =
\left\langle 
 \frac{
 \sum_{\{J_{t+1}'\}} \det B^{J'_t{}^\uparrow J'_{t+1}} \cdot
    W(\{J_t'{}^\uparrow,J_t^\downarrow\})
}{ \sum_{\{J_t\}}
      \det B^{J_t J_{t+1}{}^\uparrow}\cdot
      W(\{J_t,J_t{}^\downarrow\})}
\right\rangle_{Z_t^{\text{2-pt.},\uparrow}}
\, ,
\end{equation}
where $|J_{t+1}'| = |J_t|+1 = N_\uparrow+1$, and the chemical
potential can be calculated from the logarithmic sum of the
corresponding observables.  In the right plot of Figure \ref{fig:E0
  and mu} we show the chemical potential for a single fermion obtained
in this way. The shape of the curve depends very much on the operators
$P^\dagger, P$, i.e., the wave function used to create and annihilate
the fermion.

\bibliographystyle{JHEP}
\bibliography{tHmitcf_PoS}

\end{document}